\def\nodata{ ~$\cdots$~ }
\documentclass{aa}
\usepackage{epsfig}
\usepackage{epsf}
\usepackage{rotate}
\usepackage{rotating}
\begin{document}

   \thesaurus{08(08.06.2; 09.03.1)}

\title{The distance to the nearest star-forming clouds: MBM12 and MBM20.}

   \author{T. Hearty\inst{1}   
           \and
           M.~Fern\'andez\inst{1}
           \and
           J.M.~Alcal\'a\inst{2}
           \and
           E. Covino\inst{2}
           \and
           R. Neuh\"auser\inst{1}
           }

   \offprints{thearty@xray.mpe.mpg.de}

   \institute{Max-Planck-Inst. f\"ur Extraterrestrische Physik,
              D-85740 Garching, Germany
          \and
              Osservatorio Astronomico di Capodimonte, I-80131 Napoli, Italy}

   \date{Received; accepted}

   \maketitle

   \begin{abstract}

We present high-resolution spectra (R $\sim$ 49,000) of
stars that have parallax measurements from the
{\it Hipparcos} satellite and are projected along the line of
sight to the two nearest known star forming clouds to the Sun:
MBM12 and MBM20.  The spectra were obtained with the FOCES
Echelle Spectrograph at the 2.2 meter telescope in Calar Alto, Spain
and the wavelength range was chosen to include
the interstellar \ion{Na}{I} D lines at $\lambda 5889.950$ \AA\ and
$\lambda 5895.924$ \AA.
Since the stars are at a range of distances, we use their
spectra along with their parallaxes
from {\it Hipparcos} to determine the distance to the molecular gas.
The stars in front of the cloud do not show interstellar
\ion{Na}{I} D absorption features while the stars behind the cloud do show
interstellar absorption features.  We find that both clouds are
somewhat more distant than previously estimated.
The revised distance to MBM12 is 58$\pm$5~pc $< d <$ 90$\pm$12~pc
and the distance to MBM20 is $112\pm15$ pc $< d <$ $161\pm21$~pc.

      \keywords{Stars:formation --
                ISM:clouds
               }
   \end{abstract}

%

\section{Introduction}

The two nearest known star forming clouds to the Sun
are MBM12 and MBM20.
The MBM12 complex consists of
clouds 11, 12, and 13 from the catalog of
Magnani et al.~(1985) and is  located at (l,b) $\sim$ (159\fdg4,$-$34\fdg3)
(we will refer to the entire complex as MBM12).
It was first
identified by Lynds (1962) and appears as objects
L1453-L1454, L1457, L1458 in her catalog of dark nebulae.
The mass of the entire complex is estimated
to be $\sim$~30--200~M$_{\odot}$ based on radio maps of the region
in $^{12}$CO, $^{13}$CO and C$^{18}$O (Pound et al.~1990;
Zimmermann \& Ungerechts 1990).  
The cloud MBM20 was first identified as L1642 in the
Lynds (1962) catalog of dark nebulae. The mass of MBM20 is
$\sim$~84~M$_{\odot}$ and it is located at (l,b) $\sim$ (210\fdg9,$-$36\fdg6)
southwest of the Orion star forming complex (Magnani et al. 1985).

Both MBM12 and MBM20 are star forming clouds.
Several T~Tauri stars have been
identified in MBM12 via H$\alpha$ emission line and
X-ray surveys (Herbig \& Bell 1988; Stephenson 1986; Hearty et al. 2000).
Two binary classical T~Tauri
stars have been identified in the central region of
MBM20 (Sandell et al. 1987).
Subsequent high-resolution $^{12}$CO(J=1-0) observations
near the center of the cloud have shown that the infrared point
source IRAS04325-1419 corresponding to one of the binaries
is associated with a red and blue-shifted
bipolar outflow (Liljestr\"om et al. 1989).

Using {\it ROSAT} observations, Kuntz et al. (1997)
find evidence based on the possible detection of
0.25 keV X-ray shadows of these clouds that both are located either
within or at the edge of the Local Bubble.
They estimate that distances, $d$, of $60\leq d \leq 90$ pc to MBM12 
and $d$ $<$~100~pc to MBM20 are consistent with the foreground
0.25 keV emission seen in the direction of each cloud.  The distances are
also consistent with previous distance estimates (see Sect.~\ref{previous}).

Since MBM12 and MBM20 are both star forming clouds which are
probably located within or at the edge of the Local Bubble,
determining an accurate distance to these clouds
is important for studies of stars and of the interstellar medium.
In addition, observations of MBM12 are already scheduled
for {\it XMM} (50 ks) and {\it Chandra} (100 ks) to
investigate the diffuse X-ray emission of the Local Bubble.
Therefore we
present observations to improve the distance estimates to both clouds.
In Sect.~\ref{previous} we review previous distance estimates for these
clouds and the {\it Hipparcos} results that
revise them.  In Sect.~\ref{specsec} we present
our spectroscopic observations of {\it Hipparcos} stars along
the line of sight to both of the clouds which we use to improve
the previous distance estimates.  In Sect.~\ref{conclusions} we summarize
our investigations and suggest future observations.

\section{Previous distance estimates}
\label{previous}

The distances to MBM12 and MBM20 have previously been derived
using both photometric and spectroscopic techniques.
The spectroscopic technique used in this paper
was first used by Hobbs et al.~(1986) and Hobbs et al. (1988)
to estimate the distance to MBM12 and by Penprase (1993)
to estimate the distance to MBM20.  These studies looked for
interstellar \ion{Na}{I} D lines at $\lambda 5889.950$ \AA\ and
$\lambda 5895.924$ \AA\ in bright stars in the direction
of each cloud.  Since the distance to each star was known
via its {\it spectroscopic} parallax, the distance to the cloud
could be determined.  The stars with interstellar \ion{Na}{I} absorption
features were presumed to be behind the cloud and those without
interstellar features were presumed to be in front of the cloud
or not located behind a sufficient column density to detect the
interstellar line.  Since the {\it Hipparcos} satellite
measured the {\it trigonometric}
parallax for most of the stars used to determine the distance to
MBM12 and MBM20, it is no longer necessary
to assume a spectral type or
intrinsic luminosity (as is necessary for a spectroscopic parallax) to
measure their distance.

The stars used by Hobbs et al.~(1986),  Hobbs et al. (1988), and
Penprase (1993) to establish the distance to MBM12 and MBM20 are listed
in Table~\ref{distanceold}
with their apparent magnitude, spectral type, distance based on
spectroscopic parallax, distance based on the
{\it Hipparcos} parallax\footnote{All {\it Hipparcos} distance errors,
$\sigma_d$,
in this paper are approximated based on the parallax, $\pi$, and
standard parallax error, $\sigma_{\pi}$, from the {\it Hipparcos} catalog
using the error propagation equation
given by Bevington \& Robinson (1992).
Thus, the error in distance is given by
$\sigma_{d} = \sigma_{\pi} / \pi^{2}$, where $\sigma_d$ is in parcecs,
and $\pi$ and $\sigma_{\pi}$ are in arcseconds.}
and whether the spectrum presented in Hobbs et al.~(1986) and Hobbs et al.
(1988) showed interstellar \ion{Na}{I} absorption lines.

\subsection{MBM12}

Previous distance estimates to the cloud MBM12 have already been discussed
by Hearty et al. (2000).  We just note the {\it Hipparcos} results here.
In their search for interstellar \ion{Na}{I} D lines in the spectra of
stars along the line of sight to MBM12, Hobbs et al.~(1986) found that
the star HD18404 (distance $\sim$\ 60~pc)
showed no interstellar absorption features and is therefore presumably
in front of the cloud  and the star
HD18519/20 (distance $\sim$\ 70~pc) did show interstellar absorption features
and is therefore behind the cloud.
According to {\it Hipparcos}, the distance to HD18404 is
$\sim$~32$\pm$1~pc and the distance to HD18519/20 is $\sim$~90$\pm$12~pc.
Although the {\it Hipparcos} results indicate the distance to MBM12
is not as well constrained, it is consistent with previous
estimates (i.e., 32$\pm$1~pc $< d <$ 90$\pm$12~pc).

Hobbs et al.~(1986) note that it remains possible, although unlikely,
that a \ion{Na}{I} cloud is located in the foreground of
a more distant CO cloud at virtually the same radial velocity.
Therefore it is important to note that there are at least two stars
located behind the cloud which show a large extinction (see Fig.~1).
However, both of these stars are too faint to have been observed
with {\it Hipparcos}.  One of the stars, H0253+193, is an eclipsing
cataclysmic variable (Koyama et al. 1991; Zuckerman et al. 1992).
Zuckerman et al. (1992) estimate a distance of $\sim$ 200 pc and
a visual extinction $A_v$~$\sim$~11.5~mag based on its near infrared
colors.  In addition, Hearty et al.~(2000) showed that the optical spectrum
of another star, DC48 (Duerr \& Craine 1982), corresponds to a G9 star.
Comparing the magnitude and color measured by
Duerr \& Craine (1982) for this star ($V$ =  18.7 and V-I = 5.6~mag) to the
intrinsic values of a G9 star, it could be
a main sequence star with $A_v$ $\sim$ 8.9 mag at a distance
of $\sim$ 63~pc or a giant star with $A_v$ $\sim$ 8.4 mag at a distance
of $\sim$ 950~pc.  Future observations to better constrain the distance,
the luminosity class, and the spectral type of this star may prove that
the CO and the \ion{Na}{I} clouds along this line of sight
are one and the same and provide a new upper limit on the distance
to MBM12.

\subsection{MBM20}

The cloud MBM20 is located southwest of several cometary globules
probably associated with the Orion star forming region.
However, previous photometric and spectroscopic observations of MBM20
and the morphology of the cometary globules
suggest that MBM20 is much closer to the Sun than the cometary
globules in the region.  Studies of star formation in
Bok globules in the Gum Nebula have identified Bok globules 
with comet-like tails pointed away from the central OB
association (e.g., Reipurth 1983).
This type of large scale morphology which Bally et al. (1991)
argue is caused by the energy injected from massive stars
can also be seen in the cometary clouds surrounding the Orion OB association
where many clouds have dense heads with tails pointing away from the
center of the Orion OB association.
Whatever the mechanisms are that generate the cometary
clouds, there are no signs that MBM20 is interacting with
the bright stars in Orion (i.e., there is no cometary tail pointed away
from the OB association).

Magnani \& de Vries (1986) used
star counts to estimate an upper limit for the distance to this
cloud to be $\sim$~125~pc.  More recently, Penprase (1992) derived a photometric distance, $d$, to MBM20 of $100\leq d \leq 120$ pc.

Using the same technique we use in this paper, Penprase (1993)
arrived at an upper limit for the distance to MBM20 ($\sim$~110~pc)
by observing \ion{Na}{I} D interstellar absorption features is stars
for which a spectroscopic parallax could be determined
(cf. Hobbs et al. 1986).  The nearest star which showed
interstellar \ion{Na}{I} absorption, HD29851,
is at a distance of 110~pc based on its spectroscopic parallax.
Since none of the stars observed by Penprase (1993) were superimposed
on the cloud contours, a lower limit for the distance to the cloud could
not be determined.
However, even without a lower limit on the distance,
these observations showed that MBM20 is among the nearest
star forming molecular clouds to the Sun.
The  {\it Hipparcos} parallax for HD29851 places it at $\sim$ $161\pm21$~pc.
Therefore, the {\it Hipparcos} results indicate
the distance to the cloud could be larger than previous estimates.

\begin{table}
\label{distanceold}
\caption{Stars from Hobbs et al.~(1986), Hobbs et al. (1988), and
Penprase (1993)}
\begin{tabular}{@{}c@{\hspace{3pt}}c@{\hspace{3pt}}c@{\hspace{3pt}}c@{\hspace{2pt}}c@{\hspace{2pt}}c@{\hspace{3pt}}c@{}}
\hline
HD       & HIP      & $V$  & SpT   & \multicolumn{2}{c}{Distance [pc]}  & \ion{Na}{I}  \\
\cline{5-6}
         &          &[mag] &       & {\scriptsize spectroscopic} & {\scriptsize \it Hipparcos} &      \\
\hline 
\multicolumn{7}{c}{MBM12} \\
\hline					                        	         
18090    & \nodata  & 8.85 & F3V   & 145   & \nodata    & yes  \\
18091    & 13579    & 7.00 & A9V   & 85    & $96\pm11$  & no   \\
18190    & \nodata  & 8.98 & A9V   & 185   & \nodata    & yes  \\
18256    & 13702    & 5.63 & F6V   & 25    & $35\pm1$   & no   \\
18283    & 13723    & 8.78 & A8III & 380   & $150\pm29$ & yes  \\
18404    & 13834    & 5.80 & F5IV  & 60    & $32\pm1$   & no   \\
18484    & 13892    & 6.70 & A3III & 211   & $141\pm35$ & yes  \\
18508    & 13913    & 7.34 & F2V   & 80    & $91\pm8$   & no   \\
18519/20$^a$ & 13914    & 4.63 & A2Vs   & 70    & $90\pm12$  & yes  \\
18654    & 14021    & 6.79 & A0V   & 160   & $128\pm18$ & yes  \\
\hline 
\multicolumn{7}{c}{MBM20} \\
\hline
28497$^b$& 20922    & 5.61 & B2V & 740 & $483\pm191$ & no \\
28763    & 21110    & 6.25 & A2/A3V & 110 & $123\pm13$ & no \\
28856    & 21168    & 9.28 & A1V & 500 & $453\pm293$ & yes \\
29482    & 21577    & 7.33 & B9.5V & 210 & $441\pm437$ & yes \\
29573    & 21644    & 4.99 & A0V &  65 & $70\pm4$ & no \\
SAO149801&\nodata   & 9.76 & K5  & \nodata  & \nodata   & no \\
29851    & 21837    & 6.64 & A2IV/V & 110 & $161\pm21$ & yes \\
\hline
\end{tabular}

$^a$The angular separation listed in the {\it Hipparcos} catalog for
this pair is $1.453\pm0.002\arcsec$ and the magnitude difference is
$0.35\pm0.01$ mag. \\
$^b$Blades et al. (1997) detected 13 interstellar absorption components
in an ultra-high-resolution (R $\sim$ 938,000) \ion{Na}{I} spectrum of this
star, however, none of the components are at the radial velocity
of MBM20.

\end{table}

\begin{figure*}
\resizebox{\hsize}{!}{\includegraphics{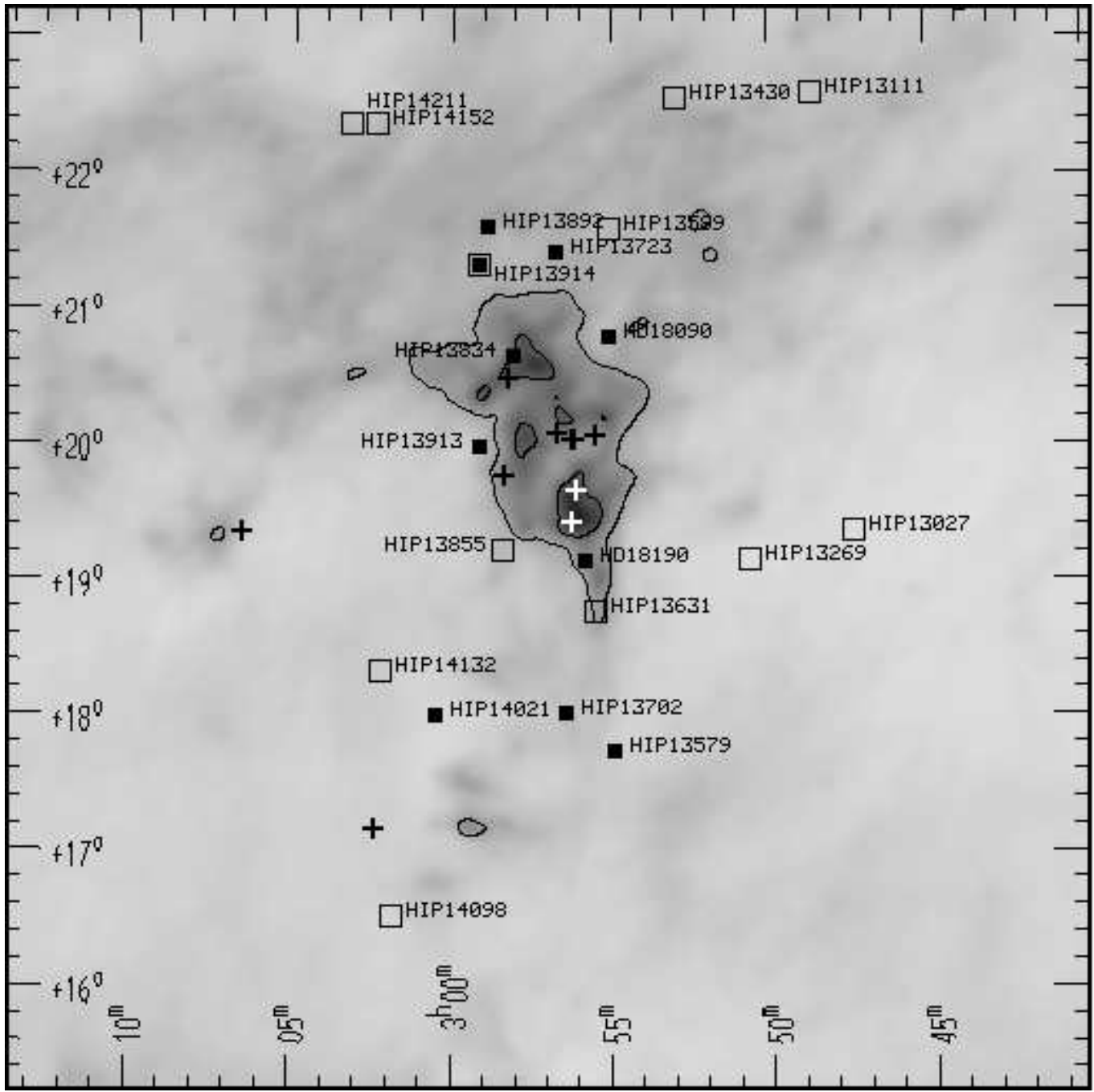}}
\caption{The gray scale image shows the extent of the IRAS 100$\mu$m emission
from MBM12.  The filled squares are the stars observed by Hobbs et al. (1986)
and Hobbs et al. (1988) to determine the distance to the cloud
and the open squares are the stars we observed to determine the
distance to the cloud.  The black plus symbols mark the locations
of the eight T Tauri stars in MBM12 (two of which
are not resolved in this figure).
The white plus symbols mark the two lines of sight toward the
cataclysmic variable H0253+193 at (RA,Dec) = (2:56:10.5,+19:26:43)
and G9 star DC48 at 
(RA,Dec) = (2:56:00.4,+19:40:44) which are known to have
$A_{\rm v}$ $>$ 5 mag (Hearty et al. 2000).
The gray scale is linear from 0 to 40 MJy sr$^{-1}$.  The contours are
at 15 MJy sr$^{-1}$ and 25 MJy sr$^{-1}$.
This figure is in J2000 coordinates.}
\label{iras12}
\end{figure*}

\begin{figure*}
\resizebox{\hsize}{!}{\includegraphics{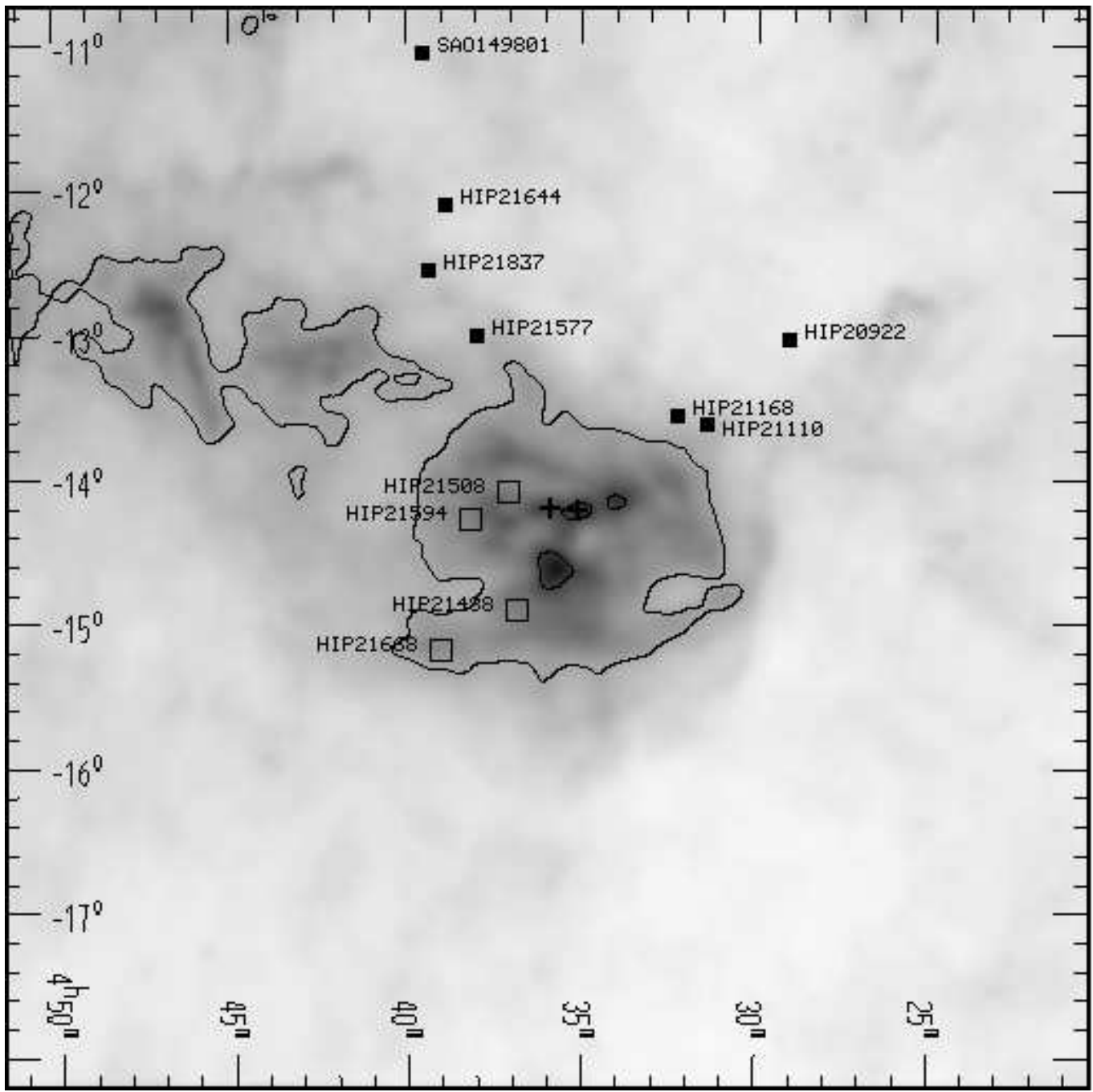}}
\caption{The gray scale image shows the extent of the IRAS 100$\mu$m emission
from MBM20.  The filled squares are the stars observed by Penprase (1992)
to determine the distance to the cloud
and the open squares are the stars we observed to determine the
distance to the cloud.  The black plus symbols mark the locations
of two binary T Tauri stars in MBM20.
The gray scale is linear from 0 to 20 MJy sr$^{-1}$.  The contours are
at 7 MJy sr$^{-1}$ and 15 MJy sr$^{-1}$.
This figure is in J2000 coordinates.}
\label{iras20}
\end{figure*}

\section{Spectroscopic Observations}
\label{specsec}
Since there are additional {\it Hipparcos} stars at intermediate distances
to the previous estimates in
the direction of both MBM12 and MBM20, we obtained high-resolution
spectroscopic observations near the \ion{Na}{I} D lines of
some of these stars to refine the distances to both clouds
(cf. Hobbs et al 1986).  The stars used by 
Hobbs et al (1986), Hobbs et al. (1988), and by us to determine the
distance to MBM12 are displayed on an IRAS 100$\mu$m image of the region
in Fig.~\ref{iras12} and the stars used by Penprase (1993) and by us
to determine the distance to MBM20 are displayed on an IRAS 100$\mu$m
image of the region around MBM20 in Fig.~\ref{iras20}

We obtained the spectra with the FOCES Echelle Spectrograph
at the 2.2-m telescope
in Calar Alto, Spain in August 1998.  The Tek$\#$13c 1K$\times$1K
chip with the FOCES Echelle Spectrograph provided a
spectral resolution, $R$ = $\lambda / \delta \lambda$ $\sim$ 49,000, derived
from measurements of the FWHM
(FWHM~$\sim$~0.12~\AA) of several well isolated emission lines of the
comparison spectra near the \ion{Na}{I} D lines.  
As a rule, the smallest measurable equivalent
width in units of m\AA\ is approximately the reciprocal dispersion in units
\AA~mm$^{-1}$ (Jaschek \& Jaschek 1987).  Therefore, given
the 24$\mu$m pixels of 
the CCD we used for the observations, the smallest measurable equivalent
width for these spectra is $\sim$ 5 m\AA.  This corresponds to a
minimum detectable column density N(\ion{Na}{I})
$\sim$ 6 $\times$ 10$^{10}$~cm$^{-2}$ in the optically thin case.
All spectra were given an initial inspection at
the telescope.  If a particular integration produced fewer
than $\sim$~1000 cts pixel$^{-1}$, at least one additional integration
was performed.

\begin{figure}
\label{specfig}
\vspace{1.0cm}
\hspace{0cm}\psfig{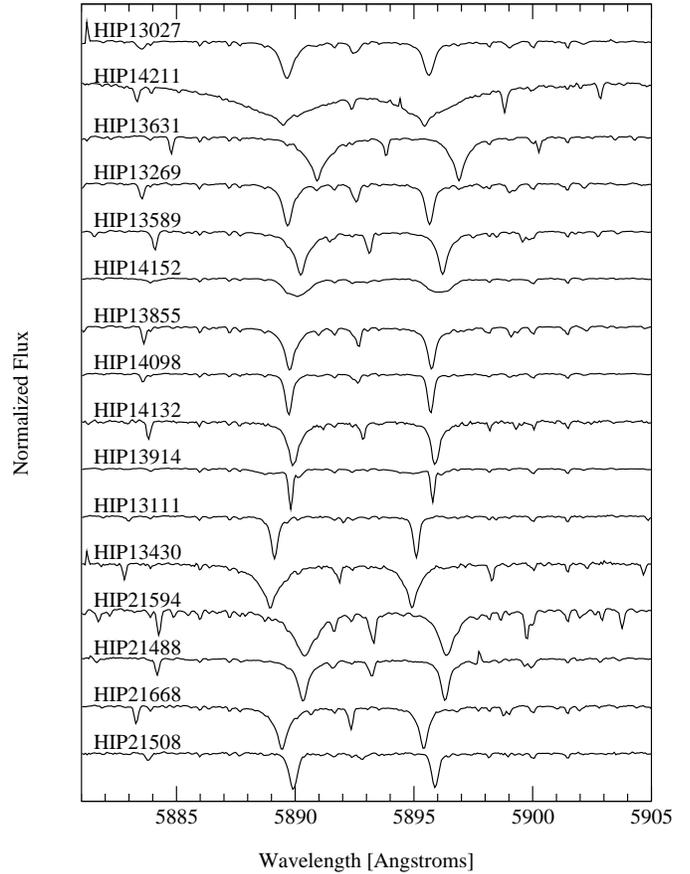}
\vspace{0cm}
\caption{Spectra are displayed for the observed Hipparcos stars
along the line of sight to MBM12 and MBM20.  The only star which shows
interstellar \ion{Na}{I} D absorption is HIP13914.  All of the
other stars have broader stellar \ion{Na}{I} D lines.}
\end{figure}

The observed stars are listed in Table~\ref{observedstars}
with their coordinates, apparent magnitude, spectral type,
{\it Hipparcos} distance, and whether the observed spectrum
contains interstellar \ion{Na}{I} features.
We observed HIP13914, although it was already
known to show interstellar
\ion{Na}{I} absorption, to ensure that we could detect the lines, if present.
Although we made no attempt to correct for telluric features,
our measured equivalent widths of the interstellar \ion{Na}{I} D1 and D2
lines 129 m\AA\ and 167 m\AA, respectively, in HIP13914 are in good
agreement with the measurements of Hobbs et al. (1986),
127 m\AA\ and 149 m\AA\, which were performed at the same spectral
resolution.  All of the observed spectra are displayed
in Fig.~\ref{specfig}.

None of the observed stars along the line of sight to MBM12
except HIP13914 showed interstellar
\ion{Na}{I} absorption, therefore the upper limit on the distance to
the cloud remains $\sim$ $90\pm12$ pc.
Since HIP13631 is the only star
we observed that is superimposed directly on the cloud
and it does not show interstellar \ion{Na}{I} absorption,
we can place a lower limit on the distance
to the cloud of $\sim$ $58\pm5$ pc.
Therefore the distance to MBM12 is $58\pm5$ pc $<$ $d$ $<$ $90\pm12$ pc.

The observed stars along the line of sight to MBM20 are all projected
on the central region of that cloud.  Since we did not detect
interstellar \ion{Na}{I} D lines in any of the observed stars,
we can place a lower limit on the distance to MBM20 of $\sim$ $112\pm15$ pc.
Therefore the distance to MBM20 is $112\pm15$ pc $<$ $d$ $<$ $161\pm21$~pc.

\begin{table}
\label{observedstars}
\caption{Observed Stars}
\begin{tabular}{@{}c@{\hspace{5pt}}c@{\hspace{5pt}}c@{\hspace{5pt}}c@{\hspace{5pt}}c@{\hspace{5pt}}c@{\hspace{5pt}}c@{}}
\hline
HIP      & RA & Dec & $V$  & SpT   & Distance  & \ion{Na}{I}  \\
         &    &      &[mag] &       & [pc]  &      \\
\hline
\multicolumn{7}{c}{MBM12} \\
\hline
13027 & 02 47 27.4 &+19 22 19 &  6.90 & G0  &$ 33\pm 2$ & no \\
14211 & 03 03 13.7 &+22 22 11 & 10.78 & M0  &$ 43\pm 4$ & no \\
13631 & 02 55 32.3 &+18 46 26 &  9.74 & K0  &$ 58\pm 5$ & no \\
13269 & 02 50 42.6 &+19 09 38 &  6.63 & F8  &$ 61\pm 3$ & no \\
13589 & 02 55 04.9 &+21 35 42 &  8.66 & F8  &$ 63\pm 5$ & no \\
14152 & 03 02 27.4 &+22 22 25 &  7.92 & F5  &$ 72\pm 6$ & no \\
13855 & 02 58 23.8 &+19 13 45 &  8.78 & G0  &$ 74\pm 8$ & no \\
14098 & 03 01 50.6 &+16 31 05 &  8.24 & F5  &$ 81\pm 8$ & no \\
14132 & 03 02 15.6 &+18 19 45 &  9.60 & G5  &$ 84\pm21$ & no \\
13914 & 02 59 12.7 &+21 20 25 &  4.63 & A2Vs&$ 90\pm12$ & yes \\
13111 & 02 48 37.4 &+22 35 54 & 10.10 & G4  &$ 91\pm13$ & no \\
13430 & 02 52 57.6 &+22 34 09 & 10.60 &\nodata&$107\pm22$ & no \\
\hline
\multicolumn{7}{c}{MBM20} \\
\hline
21594 & 04 38 10.8 &$-$14 18 14 &  3.86 &K1III&$ 34\pm 1$ & no \\
21488 & 04 36 53.9 &$-$14 55 33 &  7.93 &G2V  &$ 68\pm 7$ & no \\
21668 & 04 39 05.3 &$-$15 12 59 &  9.32 &G5   &$ 93\pm12$ & no \\
21508 & 04 37 08.3 &$-$ 14 07 03 &  8.93 &F5V  &$112\pm15$ & no \\
\hline
\end{tabular}
\end{table}

\section{Conclusions}
\label{conclusions}
We have investigated the {\it Hipparcos} parallax measurements
of stars to estimate the distance to the
two nearest star forming clouds to the Sun: MBM12 and MBM20.
The {\it Hipparcos} observations of previously observed stars
suggest that the distances to both of these
clouds are not as well constrained as previously thought, although
they are consistent with the previous estimates.  Therefore, we
obtained high resolution spectra of additional {\it Hipparcos}
stars at intermediate distances and improved the
distance estimates to both clouds.  
The distance to MBM12 is 58$\pm$5~pc $< d <$ 90$\pm$12~pc
and the distance to MBM20 is $\sim$ $112\pm15$ pc $< d <$ $161\pm21$~pc.
Since there are additional stars that were observed with
{\it Hipparcos} that are projected near each cloud (though
farther off axis than the stars we observed), future spectroscopic
observations like those presented here may further refine
our distance estimates.

\begin{acknowledgements}

We thank C. Bertout and L.M. Hobbs for comments
which helped us to improve this paper.
TH wishes to acknowledge a stipendium from the Max-Planck-Gesellschaft
and the hospitality of Professor J. Tr\"umper for support of this research.

\end{acknowledgements}


\begin{thebibliography}{}



\bibitem[]{} Bally J., Langer W.D., Wilson R.W., Stark A.A., Pound M.W., 1991,
in {\it Fragmentation of Molecular Clouds and Star Formation},
eds. E. Falgarone, F. Boulanger, \& G. Duvert (Printed in the Netherlands)


\bibitem[]{} Bevington P.R., Robinson D.K., 1992, Data Reduction and
Error Analysis for the Physical Sciences, McGraw-Hill , New York, p.\ 43


\bibitem[]{} Blades J.C., Sahu M.S., He L., et al., 1997, ApJ 478, 648








\bibitem[Duerr \& Craine 1982]{due82} Duerr R., Craine E.R., 1982, PASP 94, 567 








\bibitem[Hearty et al. 2000]{hea2000} Hearty T., Neuh\"auser R, Stelzer B.,
et al., 2000, A\&A, in press 
 
\bibitem[Hobbs et al.~(1986)]{hob86} Hobbs L.M., Blitz L.,
Magnani L., 1986, ApJ 306, L109

\bibitem[Herbig \& Bell 1988]{her88} Herbig G.H., Bell K.R., 1988, Lick
Observatory Bulletin, Santa Cruz: Lick Observatory

\bibitem[Hobbs et al. (1988)]{hob88} Hobbs L.M., Blitz L., Penprase B.E.,
Magnani L., Welty D.E., 1988, ApJ 327, 356

\bibitem[]{} Jaschek, C., Jaschek, M., 1987, The Classification of Stars,
Cambridge University Press, Cambridge, p.\ 13


\bibitem[]{} Koyama K., Takano S., Matsumoto T., et al., 1991, ApJ 377, 240

\bibitem[1997]{kun97} Kuntz K. D., Snowden S. L., Verter F., 1997, ApJ 484, 245 

\bibitem[]{} Liljestr\"om T., Mattila K., Friberg P., 1989, A\&A 210, 337

\bibitem[Lynds 1962]{lyn62} Lynds B.T., 1962, ApJS 7, 1

\bibitem[Magnani et al. 1985]{mag85} Magnani L., Blitz L., Mundy L., 1985,
ApJ 295, 402


\bibitem[Magnani \& de Vries 1986]{mag86} Magnani L., de Vries C.P., 1986,
A\&A 168, 271





\bibitem[]{} Penprase B.E., 1992, ApJS 83, 273

\bibitem[]{} Penprase B.E., 1993, ApJ 88, 433

\bibitem[Pound, Bania, \& Wilson 1990]{pou90} Pound M.W., Bania T.M.,
Wilson R.W., 1990, ApJ 351, 165



\bibitem[]{} Reipurth B., 1983, A\&A 117, 183

\bibitem[]{} Sandell G., Reipurth B., Gahm G., 1987, A\&A 181, 283

\bibitem[]{} Stephenson C.B., 1986, ApJ 300, 779




\bibitem[]{} Zimmermann T., Ungerechts H., 1990, A\&A 238, 337	

\bibitem[]{} Zuckerman B., Becklin E.E., McLean I.S., Patterson J., 1992,
ApJ 400, 665

\end{thebibliography}
\end{document}